# Object Oriented Model for Evaluation of On-Chip Networks

[1]Sheraz Anjum, [1]Ehsan Ullah Munir, [2]Waqas Anwar and [3]Nadeem Javaid
[1]COMSATS Institute of Information Technology, Wah Cantt, Pakistan
[2]COMSATS Institute of Information Technology, Abbottabad, Pakistan
[3]COMSATS Institute of Information Technology, Islamabad, Pakistan

**Abstract:** The Network on Chip (NoC) paradigm is rapidly replacing bus based System on Chip (SoC) designs due to their inherent disadvantages such as non-scalability, saturation and congestion. Currently very few tools are available for the simulation and evaluation of on-chip architectures. This study proposes a generic object oriented model for performance evaluation of on-chip interconnect architectures and algorithms. The generic nature of the proposed model can help the researchers in evaluation of any kind of on-chip switching networks. The model was applied on 2D-Mesh and 2D-Diagonal-Mesh on-chip switching networks for verification and selection of best out of both the analyzed architectures. The results show the superiority of 2D-Diagonal-Mesh over 2D-Mesh in terms of average packet delay.

## INTRODUCTION

According to International Technology Roadmap for Semiconductors (International Technology Roadmap for Semiconductors, 2004), the process technologies would continue to scale down. Current bus based System on Chip (SoC) designs could not scale with the advancement of process technologies and is becoming a bottleneck due to contention and congestion issues. More over global wire delays, reusability, less modularity and scalability issues have added to the problems of bus based SoC designs. To fill the gap created due to the inherent disadvantages of bus-based SoCs, more modular and scalable design methodologies (Benini and Micheli, 2002; Shashi *et al*., 2002; Axel and Hannu, 2003) known as Network on Chip (NoC) have been proposed in the literature. The use of globally asynchronous and locally synchronous concept in NoCs (Hemani *et al*., 1999) has disintegrated the design of resources from the rest of the network.

Design and selection of appropriate architecture for on-chip communication has a key role in the design and implementation of the complete platform for NoC. Authors in Faraydon *et al*. (2001), Alireza *et al*. (2005), Partha *et al*. (2005), Hemayet *et al*. (2005), Sheraz *et al*. (2011, 2009) and Yi *et al*. (2002) have proposed, evaluated, or analyzed different on-chip architectures. The selection of best out of the proposed or designed NoC architecture requires intelligent tools. Keeping in view of the fact that currently very few tools are available for simulation and performance evaluation of NoCs, this study presents a generic object oriented model under NS-2 (The NS Manual, 2006; Simulator for Beginners, 2003) for the simulation and evaluation of any kind of on-chip switching network.

**Proposed simulation model:** Any NoC network or proposal could comprise of the following basic entities:

- Resources/nodes
- Links/wires
- Switches/routers
- Switching algorithm

The switches and resources can be connected through links to form a switching and live on-chip interconnection network called NoC. The sending resource can use different paths through the switches to send its messages to the desired destination. These messages can be considered as traffic between this sender/target pair. The switch acts as a by passer of the message from its input to one of its output. It accomplishes this work depending on an algorithm known as routing or switching algorithm. To simulate and evaluate any NoC architecture, we need all the basic entities listed above as well as traffic model and some kind of queue monitoring for the purpose of analysis and performance evaluation of one entity or the complete platform. The careful study of NS-2 (The NS Manual, 2006; Simulator for Beginners, 2003) leads us to know that these required entities are readily available in the form of objects. These objects can be accessed by writing an OTCL script that acts as an interface to the C++ object library. Therefore, depending on these objects, we developed an object oriented simulation model shown in Fig. 1.

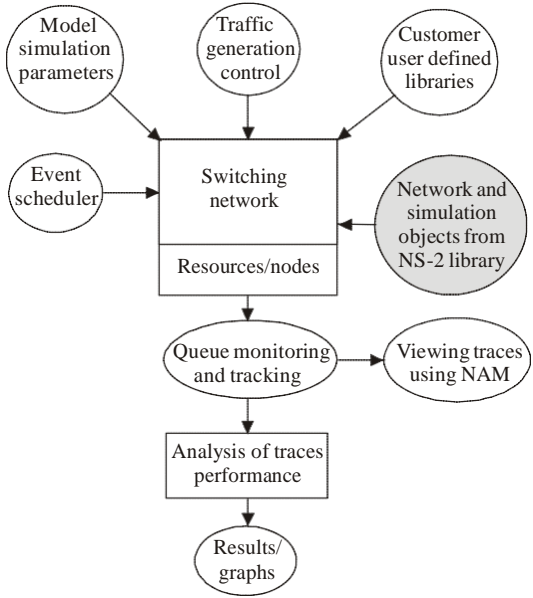

Fig. 1: Proposed object oriented model for NoC simulation

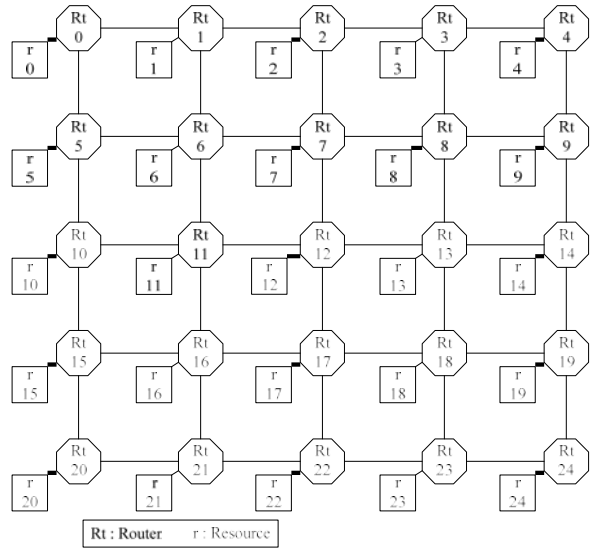

Fig. 2: 2D-mesh switching network

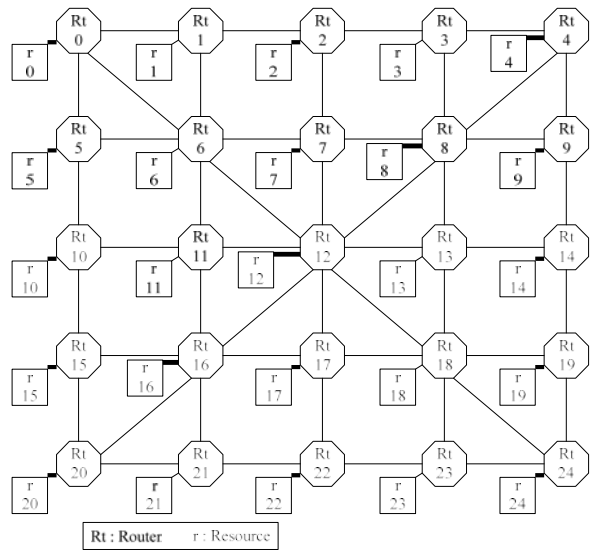

Fig. 3: 2D-diagonal-mesh switching network

In this model, the blocks with filled color show that these objects are available in NS-2 while white blocks show that the objects or activities have to be accomplished by the user. The resources, links, queue monitoring and visualization of the network is available in NS-2. In addition to the library of objects from NS-2, user can develop and use his/her own library or a custom library that must be embedded into the NS-2 environment according to its internal structure.

If a user needs to evaluate any NoC platform using this model then he/she has to perform following functions:

- Define his/her switching network by borrowing components from NS-2 or user defined libraries.
- Develop model for the selection of source/destination pairs that is traffic scenario generation depending on the communication requirement of the network under consideration.
- Has to model important parameters such as bandwidths, delays and packets size specific to the NoC platform under consideration.
- Schedule the events between the communicating resources.

At the end the generated trace files must be used for the analysis and performance evaluation of the architecture under consideration.

**Simulation environments:** In order to apply the proposed simulation model we have to use different switching networks and traffic models for source/destination pairs. The interconnect architectures used for simulation are discussed in Sheraz *et al*. (2009) and are known as 2D-Mesh and 2D-Diagonal-Mesh. These on-chip interconnect architectures are shown in Fig. 2 and 3.

Three traffic selection models were used for comparison of the switching networks of Fig. 2 and 3. Figure 4 reveals the details of the first and second models. The major difference in the 2D and 2D-

Diagonal traffic models is the selection of neighbors and non-neighbors according to 2D-Mesh and 2D-Diagonal-Mesh architectures respectively i.e., in 2D traffic model resources at the corner have only two neighbors while in 2D-Diagonal they have three neighbors.

Similar is the case in the selection of non-neighbors. In both the models we have to set 'Range1' between 0 and 1. If we set it to 0, it means the algorithm will always select a random non-neighbor of the source Si, whereas on the opposite end if we set it to 1, it means a random neighbor of source Si will always be selected. The middle values of 'Range1' will change the probability of selection between neighbors and non-neighbors of source Si.

The third selection model given by Eq. (1) is specifically modeled to check the effectiveness of 2D-Diagonal-Mesh over 2D-Mesh network:

$$D_i = (N-1) - S_i \qquad (1)$$

where,
'Di' : The i$^{th}$ destination
'Si' : The sources respectively
'N' : The total number of resources used

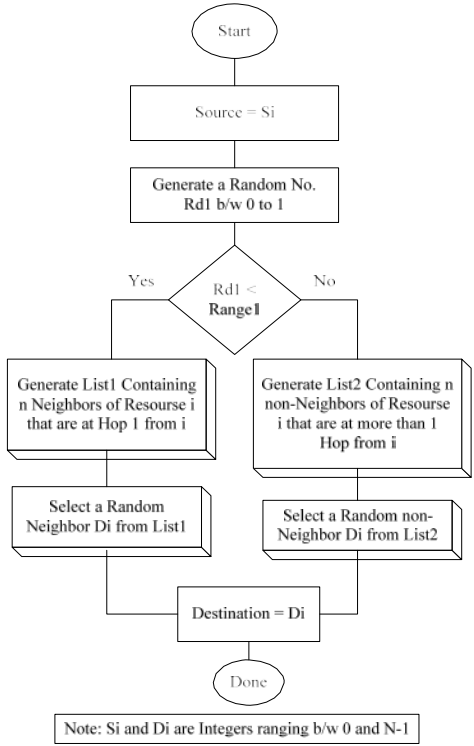

Fig. 4: 2D and 2D-diagonal traffic selection models

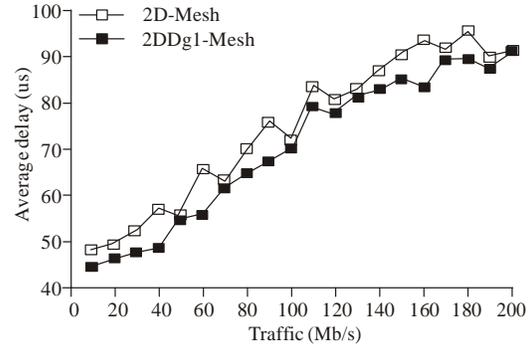

Fig. 5: Delay vs traffic rate using 2D random traffic model

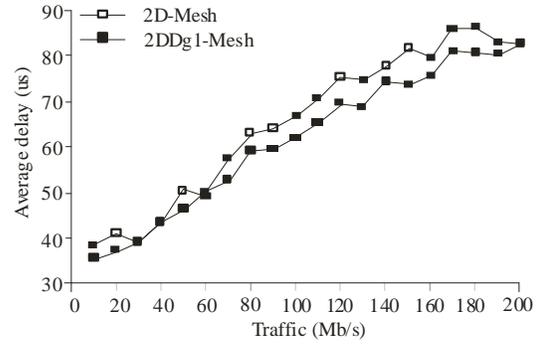

Fig. 6: Delay vs traffic using 2D-diagonal 75% local traffic model

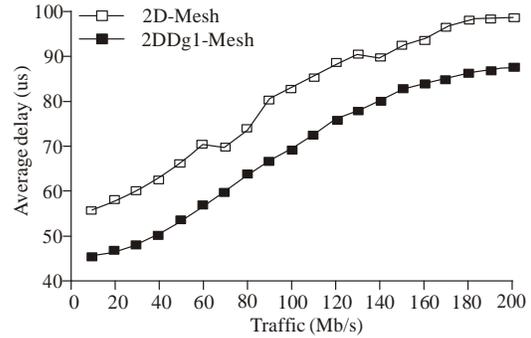

Fig. 7: Delay vs traffic rate using fixed traffic model

## RESULTS OF COMPARISON

The switching networks of Fig. 2 and 3 were evaluated for average packet delay at targets using the traffic selection models of Fig. 4 and Eq. (1). The average delay of packets is given in Eq. (2):

$$D_a = \sum_{i=0}^{P} D_{Li} / P \qquad (2)$$

where,
$D_a$ = Average Packet Delay
$D_{Li}$ = End to end delay of packet i
P = Total packets generated in one simulation

Figure 5 and 6 shows the average packet delay Vs traffic rate comparison for both the networks using 2D and 2D-Diagonal traffic models respectively. It is evident from the graphs that 2D-Diagonal-Mesh almost

Figure 7 shows that under the fixed traffic selection model 2D-Diagonal-Mesh has even lesser average delay than 2D-Mesh. These results guided us to select 2D-Diagonal-Mesh switching network for implementation of our NoC platform in this particular scenario.

## CONCLUSION

This study tried to fill the gap created due to the lack of tools available for the simulation and performance evaluation of on-chip communication networks. The study proposed an object oriented model for the simulation and evaluation of any kind of on-chip switching networks which uses the publicly available network simulator NS-2 The NS manual (2006) and Simulator for Beginners (2003). The model was applied on 2D-Mesh and 2D-Diagonal-Mesh on-chip interconnects architectures for verification. Results show the ease of selection out of many different kinds of on-chip switching networks. Hence the proposed model can help NoC researchers to select best out of many different on-chip switching networks, that in turn speedup the design cycle of Network on Chip.

## ACKNOWLEDGMENT

This study is sponsored by COMSATS Institute of Information Technology, Pakistan.

## REFERENCES

Alireza, V., T. Ahmadreza and H.F. Mohammad, 2005. Hierarchical graph: A new cost effective architecture for network on chip. Proceedings of the 2005 International Conference on Embedded and Ubiquitous Computing, (EUC'05), Springer-Verlag Berlin, Heidelberg, pp: 311-320.

Axel, J. and T. Hannu, 2003. Networks on Chip. Kluwer Academic Pub., Stockholm, NY, pp: 85-106.

Benini, L. and G. D. Micheli, 2002. Networks on chip: A new SoC paradigm. Proc. IEEE Comput., 35(1): 70-78.

Faraydon, K., N. Anh, D. Sujit and R. Ramesh, 2001. On-chip communication architecture for OC-768 Network Processors. Proceedings of the 38th annual Design Automation Conference, (DAC '01), ACM New York, USA, pp: 678-683.

Hemani, A., T. Meincke, S. Kumar, A. Postula, T. Olsson, *et al.*, 1999. Lowering power consumption in clock by using globally asynchronous locally synchronous design style. Automation Conference, (DAC '99), ACM New York, USA, pp: 873-878.

Hemayet, H., M.A. Muhammad and M.I. Muhammad, 2005. Extended-butterfly FAT Tree Interconnection (EFTI) architecture for NoC. IEEE Pacific Rim Conference on Communications, Computers and signal Processing, 24-26 Aug., Dept. of Comput. Sci. and Eng., Bangladesh Univ. of Eng. and Technol., Dhaka, Bangladesh, pp: 613-616.

International Technology Roadmap for Semiconductors, 2004. World Semiconductor Council, Edition.

Partha, P.P., G. Cristian, J. Michael, I. André and S. Resve, 2005. Performance evaluation and design trade-offs for network- on-chip interconnect architectures. Proc. IEEE Trans. Comput., 54(8): 1025-1040.

Shashi, K., Jantsch, M. Millberg, J. Öberg and A. Hemani, 2002. A network on chip architecture and design methodology. ISVLSI 2000: IEEE Computer Society Annual Symposium on Vlsi-New Paradigms for Vlsi Systems Design, IEEE Computer Soc., Los Alamitos, pp: 117-124.

Sheraz, A., J. Chen, P.P. Yue and L. Jian, 2009. A delay optimized architecture for on-chip communication. J. Electr. Sci. Techn. China, 7(2): 104-109.

Sheraz, A., U.M. Ehsan and W.N. Muhammad, 2011. Simulation and performance evaluation of network on chip architectures and algorithms using CINSIM. J. Basic Appl. Sci. Res., 1(10): 1594-1602.

Simulator for Beginners, 2003. Lecture Notes. Univ. de Los Andes, Merida, Venezuela, Sophia-Antipolis, France, Ch: 2-5.

The NS Manual, 2006. The VINT Project. UC Berkeley, LBL, USC/ISI and Xerox PARC, Ch: 5-10.

Yi-Ran, S., K. Shashi, J. Axel, 2002. Simulation and evaluation for a network on chip architecture using NS-2. Proceedings of 20th NORCHIP Conference.